\documentclass[amsfonts,amssymb,amsmath,byrevtex,twoside]{revtex4}
\usepackage{hyperref}
\newtheorem{theorem}{Theorem}
\begin{document}

\title{Statistical models of hadron production -- simple models for complicated
processes}
\thanks{Invited lecture at the Helmoltz
International Summer School \emph{''Dense Matter in Heavy Ion Collisions and
Astrophysics''}, JINR, Dubna, 21.08--1.09.2006}
\author{Ludwik Turko}
\email{turko@ift.uni.wroc.pl}
\affiliation{Institute of Theoretical Physics, University of Wroclaw,\\
Pl. Maksa Borna 9, 50-204  Wroc{\l}aw, Poland}
\date{March 19, 2007}

\begin{abstract}
Thermal statistical models are simple and effective tool to describe particle
production in high energy heavy ion collision. It is shown that for higher moments
finite volume corrections become important observable quantities. They make possible
to differentiate between different statistical ensembles even in the thermodynamic
limit.
\end{abstract}
\maketitle

\section{Introduction}

Particle production yields are nicely reproduced by thermal models, based on the
assumption of noninteracting gas of hadronic resonances \cite{brs}. This simplicity
can be however misleading as particle yields and particle ratios are not very
sensitive to the underlying model. The main ingredient of the statistical models are
probability densities, which allow to extract the whole physical information. The
only way to reproduce those probability distributions is by means of higher and
higher probability moments. These moments are in fact the only quantities which are
phenomenologically available and can be used for the verification of theoretical
predictions.

Particle yields in heavy ion collision are the first moments, so they lead to rather
crude comparisons with the model. Fluctuations and correlations are second moments
so they allow for the better understanding of physical processes in the thermal
equilibrium.

Fluctuations and correlations measured in heavy ion collision processes give better
insight into dynamical and kinematical properties of the dense hadronic medium
created in ultrarelativistic heavy ion collisions.  Systems under considerations are
in fact so close to the thermodynamic limit that final volume effects seem to be
unimportant --- at least when productions yields are considered.

The aim of the paper is to show that finite volume effects become more and more
important when higher moments, \emph{e.g.} correlations and fluctuations are
considered.

A preliminary analysis of the increasing volume effects was given in
\cite{crt1,crt2}. It has been rigorously shown an influence of $\mathcal{O}(1/V)$
terms for a new class physical observables --- semi-intensive quantities
\cite{crt2}. Those results completely explained also ambiguities noted in
\cite{begun}, related to ''spurious non-equivalence'' of different statistical
ensembles used in the description of heavy ion collision processes.

We start from the simple example of the standard statistical physics. Is is shown
that even in that case a notion of semi--intensive quantities is relevant for the
physical situation. In the next step we consider an abelian symmetry corresponding
to one conserved charge.

\section{Choice of variables}

In the thermodynamical limit the relevant probabilities distributions are those
related to densities. These distributions are expressed by moments calculated for
densities --- not for particles. In the practice, however, we measure particles ---
not densities as we do not know related volumes. Fortunately, volumes can be omitted
by taking corresponding ratios.

Let us consider \emph{e.g.} the density variance $\Delta n^2$. This can be written
as
\[\Delta n^2=\langle n^2\rangle - \langle n\rangle^2 =
\frac{\langle N^2\rangle - \langle N\rangle^2}{V^2}\,.\]%
By taking the relative variance
 \[\frac{\Delta n^2}{\langle n\rangle^2}=\frac{\langle N^2\rangle - \langle N\rangle^2}{\langle N\rangle^2}\,,\]
 volume-dependence vanishes.

\subsection{Semi-intensive variables}

A special care should be taken for calculations of ratios of particles moments.
Although moments are extensive variables their ratios can be finite in the
thermodynamic limit. These ratios are examples of semi-intensive variables. They are
finite in the thermodynamic limit but those limits depend on volume terms in density
probability distributions. One can say that semi-intensive variables ''keep memory''
where the thermodynamic limit is realized from.

Let consider as an example the scaled  particle variance
\[\frac{\langle N^2\rangle - \langle N\rangle^2}{\langle N\rangle}=
V\frac{\langle n^2\rangle - \langle n\rangle^2}{\langle n\rangle}\,.\]%

The term
\[\frac{\langle n^2\rangle - \langle n\rangle^2}{\langle n\rangle}\,.\]
tends to zero in the thermodynamic limit as $\mathcal{O}(V^{-1})$. So a behavior of
the scaled particle variance depends on the $\mathcal{O}(V^{-1})$ term in the scaled
density variance. A more detailed analysis of semi-intensive variables is given in
\cite{crt2}.

To clarify this approach let us consider a well known classical problem of Poisson
distribution but taken in the thermodynamic limit.

\section{Grand canonical and canonical ensembles}
\subsection{Poisson distribution in the thermodynamic limit}

Let us consider the grand canonical ensemble of noninteracting gas. A corresponding
statistical operator is
  \begin{equation}\label{stat operator GC-P}
    \hat{D}=\frac{\,e^{-\beta\hat H+\gamma\hat N}}{\,\text{Tr}{\,e^{-\beta\hat H+\gamma\hat N}}}
\end{equation}

This leads to the partition function
\begin{equation}\label{GC-P part fn}
    \mathcal{Z}(V,T,\gamma)=\,e^{z\,e^\gamma}\,.
\end{equation}

where $z$ is  one-particle partition function
\begin{equation}
z(T,V)=\frac{V}{(2\pi)^3}\int d^3p\,\,e^{-\beta E(p)} \equiv V z_0(T)\,,
\end{equation}

A $\gamma$ parameter ($=\beta\mu$) is such to provide the given value of the average
particle number \mbox{$\langle N\rangle=V\langle n\rangle$}. This means that
\begin{equation}\label{particle factor}
    \,e^{\gamma}= \frac{\langle n\rangle}{z_0}\,.
\end{equation}

Particle moments can be written as
\begin{equation}\label{particle moments}
    \langle N^k\rangle= \frac{1}{\mathcal{Z}}\frac{\partial^k\mathcal{Z}}
    {\partial\gamma^k}\,.
\end{equation}

The parameter $\gamma$ is  taken in final formulae as a function $\gamma(\langle
n\rangle,z_0)$ from Eq \eqref{particle factor}.

The resulting probability distribution to obtain $N$ particles under condition that
the average number of particles is $\langle N\rangle$ is equal to Poisson
distribution

\[P_{\langle N\rangle}(N)=\frac{{\langle N\rangle}^N}{N!}\,e^{-\langle N\rangle}\,.\]

We introduce corresponding probability distribution $\mathcal{P}$ for the particle
number density $n=N/V$

\begin{equation}\label{probab dens}
\mathcal{P}_{\langle n\rangle}(n;V)= V P_{V\langle n\rangle}(V n)=V\frac{(V\langle
n\rangle)^{V n}}{\Gamma(V n+1)} \,e^{-V\langle n\rangle}\,.
\end{equation}

 For large $V n$  we are using an asymptotic form of Gamma function

 \[\Gamma(V n+1)\sim\sqrt{2\pi}(V n)^{V n-1/2}\,e^{-V n}\left\{1+\frac{1}{12 V n}+
 \mathcal{O}(V^{-2})\right\}\,.\]

This gives

\begin{equation}\label{prob dens as1}
    \mathcal{P}_{\langle n\rangle}( n;V)\sim V^{1/2}\frac{1}{\sqrt{2\pi n}}
\left(\frac{\langle n\rangle}{ n}\right)^{V n} \,e^{V( n-\langle
n\rangle)}\left\{1-\frac{1}{12 V n}+\mathcal{O}(V^{-2})\right\}
\end{equation}

This expression in singular in the $V\to\infty$ limit. To estimate a large volume
behavior of the probability distribution \eqref{probab dens} one should take into
account a generalized function limit. So we are going to calculate an expression

 \[\langle G\rangle_V=\int dn\, G( n)\mathcal{P}_{\langle n\rangle}(n;V)\,,\]

where $\mathcal{P}_{\langle n\rangle}(n;V)$ is replaced by the asymptotic form from
Eq \eqref{prob dens as1}. In the next to leading order in $1/V$ one should calculate

\begin{equation}\label{Poiss t lim}
    V^{1/2}\frac{1}{\sqrt{2\pi}}\int d n\frac{G( n)}{ n^{1/2}}\,e^{V S( n)} -
V^{-1/2}\frac{1}{12\sqrt{2\pi}}\int d n\frac{G( n)}{ n^{3/2}}\,e^{V S( n)}\,.
\end{equation}

where
\[S( n)= n\ln\langle n\rangle -  n\ln n +  n - \langle n\rangle\,.\]

An asymptotic expansion of the function $\langle G\rangle_V$ is given by the
classical Watson-Laplace theorem

\begin{theorem}
  Let $I=[a,b]$ be the finite interval such that
  \begin{enumerate}
    \item $\max\limits_{x\in I} S(x)$ is reached in the single point $x=x_0$, \mbox{$a<x_0<b$}.
    \item $f(x),S(x)\in C(I)$.
    \item $f(x), S(x)\in C^\infty$ in the vicinity of $x_0$, and $S^{''}(x_0)\neq 0$.
  \end{enumerate}
  Then, for $\lambda\to\infty,\ \lambda\in S_\epsilon$, there is an asymptotic expansion
\begin{subequations}\label{laplace}
  \begin{eqnarray}
    F[\lambda]&\thicksim &\,e^{\lambda S(x_0)}\sum\limits_{k=0}^\infty c_k\lambda^{-k-1/2}\,,
    \label{laplace main}\\
    c_k &=&\frac{\Gamma(k+1/2)}{(2k)!}\left(\frac{d}{dx}\right)^{2k}
\left.\left[f(x)\left(\frac{S(x_0)-S(x)}{(x-x_0)^2}\right)^{-k-1/2}\right]\right\vert_{x=x_0}\,.
\label{laplace coeff}
\end{eqnarray}
\end{subequations}
  $S_\epsilon$~is here a segment $|\arg z|\leqslant\frac{\pi}{2}-\epsilon<\frac{\pi}{2}$ in
the complex $z$-plane.
\end{theorem}

To obtain $\mathcal{O}(1/V)$ formula the first term in \eqref{Poiss t lim} should be
calculated till the next to leading order term in $1/V$. For the second term it is
enough to perform calculations in the leading order only.

The first term gives the contribution
\begin{subequations}
  \begin{equation}\label{first}
  V^{1/2}\frac{1}{\sqrt{2\pi}}\int d n\frac{G( n)}{ n^{1/2}}\,e^{V S( n)}
   = G(\langle n\rangle)+\frac{1}{12\langle n\rangle V} G(\langle n\rangle)+\frac{\langle n\rangle}{2
    V}G^{''}(\langle n\rangle)\,,
\end{equation}
and the second term gives
  \begin{equation}\label{second}
   V^{-1/2}\frac{1}{12\sqrt{2\pi}}\int d n\frac{G( n)}{ n^{3/2}}\,e^{V S( n)} =
    \frac{1}{12\langle n\rangle V} G(\langle n\rangle)\,,
\end{equation}
\end{subequations}
So we have eventually
\begin{equation}\label{t lim 2}
    \langle G\rangle_V = G(\langle n\rangle) + \frac{\langle n\rangle}{2V}G^{''}(\langle n\rangle) +
    \mathcal{O}(V^{-2})\,,
\end{equation}
for any function $G$.

This gives us the exact expression for the density distribution \eqref{probab dens}
in the large volume limit
\begin{equation}\label{poison t lim 2}
    \mathcal{P}_{\langle n\rangle}(n;V)\sim\delta( n-\langle n\rangle)+\frac{\langle n\rangle}{2
    V}\,\delta^{''}( n-\langle n\rangle)+\mathcal{O}(V^{-2})\,.
\end{equation}

We are now able to obtain arbitrary density moments up to $\mathcal{O}(V^{-2})$
terms.
\begin{equation}\label{moments}
    \langle n^k \rangle_V = \int dn\, n^k \mathcal{P}_{\langle
    n\rangle}(n;V) = \langle n\rangle^k +\frac{k(k-1)}{2V}\langle
    n\rangle^{k-1}+\mathcal{O}(V^{-2})\,.
\end{equation}

We have for the second moment (intensive variable!)

\[\langle n^2 \rangle_V = \langle n\rangle^2 + \frac{\langle n\rangle}{V}+\mathcal{O}(V^{-2})\,.\]

This means
\begin{equation}\label{density limit}
    \Delta n^2=\frac{\langle n\rangle}{V}\to 0\,.
\end{equation}
as expected in the thermodynamic limit.

The particle number and its density are fixed in the canonical ensemble so
corresponding variances are always equal to zero. The result \eqref{density limit}
can be seen as an example of the equivalence of the canonical and grand canonical
distribution in the thermodynamic limit. This equivalence is clearly visible from
the Eq \eqref{poison t lim 2} where the delta function in the first term can be
considered as the particle number density distribution in the canonical ensemble.

A more involved situation appears for particle number moments (extensive variable!).
Eq \eqref{moments} translated to the particle number gives
\begin{equation}\label{particle moments 2}
        \langle N^k\rangle = V^k\langle n\rangle^k + V^{k-1}\frac{k(k-1)}{2}\langle
        n\rangle^{k-1}+\mathcal{O}(V^{k-2})\,,
\end{equation}

 One gets for the scaled variance (semi-intensive variable!)
\begin{equation}\label{scaled variance}
    \frac{\Delta N^2}{\langle N\rangle}=1\,,
\end{equation}
what should be compared with zero obtained for the canonical distribution.

The mechanism for such a seemingly unexpected behavior is quite obvious. The grand
canonical and the canonical density probability distributions tend to the same
thermodynamic limit. There are different however for any finite volume.
Semi-intensive variables depend on coefficients at those finite volume terms so they
are different also in the thermodynamic limit.

\subsection{Energy distribution}
It is interesting to perform similar calculation for the energy distribution in both
ensembles. Energy moments and an average energy density can be written as
\begin{equation}\label{energy moments}
    \langle E^k\rangle = (-1)^k\frac{1}{\mathcal{Z}}\frac{\partial^k\mathcal{Z}}
    {\partial\beta^k}\,;\qquad \langle \epsilon\rangle=-\frac{d z_0}{d\beta}\,e^\gamma\,.
\end{equation}

One gets from Eq \eqref{energy moments}
\begin{equation}\label{en moments GC-P}
    \langle E^k\rangle= V^k\langle\epsilon\rangle^k+V^{k-1}\frac{k(k-1)}{2}\langle\epsilon\rangle^{k-2}
    \frac{\langle n\rangle}{z_0}\frac{d^2 z_0}{d\beta^2}+\mathcal{O}(V^{k-2})\,.
\end{equation}
The grand canonical energy density distribution follows
\begin{equation}\label{energ probab}
    \mathbf{P}(\epsilon|\langle  n\rangle,\langle\epsilon\rangle)=
    \delta\left(\epsilon-\langle\epsilon\rangle\right)+
    \frac{\langle n\rangle}{2 V}\,
    \mathcal{R}^{GC}\left(\frac{\langle\epsilon\rangle}{\langle n\rangle}\right)
    \delta^{''}(\epsilon -\langle \epsilon\rangle)
    +\mathcal{O}(V^{-2})\,.
\end{equation}
$\mathcal{R}^{GC}$ is given here as
\[\mathcal{R}^{GC}\left(\frac{\langle\epsilon\rangle}{\langle n\rangle}\right)=
\left.\frac{1}{z_0}\frac{d^2
z_0}{d\beta^2}\right|_{\beta=\beta(\langle\epsilon\rangle/\langle n\rangle)}\,.\]

For the canonical distribution a corresponding statistical operator is
  \begin{equation}\label{stat operator C-P}
    \hat{D}=\frac{\,e^{-\beta\hat H}}{\,\text{Tr}{\,e^{-\beta\hat H}}}
\end{equation}

This leads to the partition function
\begin{equation}\label{C-P part fn}
    \mathcal{Z}(V,T)=\frac{z^N}{N!}=\frac{\,e^{Vn\log z}}{N!}\,.
\end{equation}

Internal energy moments are given by Eq \eqref{energy moments}. In particular
\begin{equation}\label{av energy C-P}
    \langle \epsilon\rangle=-\frac{n}{z_0}\frac{d z_0}{d\beta}\,.
\end{equation}

For the energy moments one gets now
\begin{equation}\label{en moments C-P}
    \langle E^k\rangle=V^k\langle \epsilon\rangle^k +
    V^{k-1}\frac{k(k-1)}{2}\langle
    \epsilon\rangle^{k-2}n\frac{\partial}{\partial\beta}\left(\frac{1}{z_0}
    \frac{\partial z_0}{\partial\beta}\right)+\mathcal{O}(V^{k-2})\,.
\end{equation}

A corresponding probability distribution is
\begin{equation}\label{energ probab C-P}
    \mathbf{P}(\epsilon|n,\langle\epsilon\rangle)=
    \delta\left(\epsilon-\langle\epsilon\rangle\right)+
    \frac{n}{2 V}\,
    \mathcal{R}^{C}\left(\frac{\langle\epsilon\rangle}{n}\right)
    \delta^{''}(\epsilon -\langle \epsilon\rangle)
    +\mathcal{O}(V^{-2})\,,
\end{equation}
where $\mathcal{R}^{C}$ is given here as
\[\mathcal{R}^{C}\left(\frac{\langle\epsilon\rangle}{n}\right)=
\left.\frac{\partial}{\partial\beta}\left(\frac{1}{z_0}
    \frac{\partial z_0}{\partial\beta}\right)
    \right|_{\beta=\beta(\langle\epsilon\rangle/n)}\,.\]

\section{High energy statistical physics}

Although the spirit and the philosophy of the statistical approach remains the same,
ingredients of statistical models used in high energy problems are different. The
main difference is that a  number of particles is not longer conserved so we have no
chemical potentials related to that quantity. The only nontrivial chemical
potentials are those related to conserved charges, so the role of internal
symmetries is a crucial one. For abelian charges, as electric charge or baryonic
charge an introduction of corresponding potential is rather obvious -- it is just a
Lagrange multiplier at a generator of the $U(1)$ symmetry. For non-abelian
symmetries, as \emph{e.g.} for the isotopic $SU(2)$ symmetry, problem is more
involved. It appears \cite{TurRed} that in such a case the only relevant chemical
potentials are those related to so called Cartan subgroup -- a maximal abelian
subgroup of the given non-abelian group of internal symmetry.

So for the hot hadronic gas a well approximated internal symmetry is $SU(3)$ or
$SU(4)$ with a charm taken into account. On the top of it we have the exact $U(1)$
baryon number conservation. $SU(3)$ symmetry leads to two chemical potentials,
related to the third isospin component and to the hypercharge. Those two chemical
potentials are supplemented by the baryonic potential. One uses for practical
reasons another set of chemical potentials, which are linear combinations of the
basic set. These are the electric charge and the strangeness chemical potentials
together with the unchanged baryonic potential.

Then, for the simplest case of an ideal hadron gas in thermal and chemical
equilibrium, which consists of $l$ species of particles, energy density $\epsilon$,
baryon number density $n_{B}$, strangeness density $n_{S}$ and electric charge
density $n_(Q)$ read ($\hbar=c=1$ always)
\begin{subequations}\label{eqstate}
  \begin{align}
\epsilon = { 1 \over {2\pi^{2}}} \sum_{i=1}^{l} (2s_{i}+1)
\int\limits_{0}^{\infty}dp\,{ { p^{2}E_{i} } \over { \exp \left\{ {{ E_{i} - \mu_{i}
} \over T} \right\} + g_{i} } }
&\ ,\\
n_{B}={ 1 \over {2\pi^{2}}} \sum_{i=1}^{l} (2s_{i}+1) \int\limits_{0}^{\infty}dp\,{
{ p^{2}B_{i} } \over { \exp \left\{ {{ E_{i} - \mu_{i} } \over T} \right\} + g_{i} }
}
&\ ,\\
n_{S}={1 \over {2\pi^{2}}} \sum_{i=1}^{l} (2s_{i}+1) \int\limits_{0}^{\infty}dp\,{ {
p^{2}S_{i} } \over { \exp \left\{ {{ E_{i} - \mu_{i} } \over T} \right\} + g_{i} } }
&\ ,\\
n_Q={1 \over {2\pi^{2}}} \sum_{i=1}^{l} (2s_{i}+1) \int\limits_{0}^{\infty}dp\,{ {
p^{2}Q_{i} } \over { \exp \left\{ {{ E_{i} - \mu_{i} } \over T} \right\} + g_{i} } }
&\ .
\end{align}
\end{subequations}
where $E_{i}= ( m_{i}^{2} + p^{2} )^{1/2}$ and $m_{i}$, $B_{i}$, $S_{i}$, $\mu_{i}$,
$s_{i}$ and $g_{i}$ are the mass, baryon number, strangeness, chemical potential,
spin and a statistical factor of specie $i$ respectively (we treat an antiparticle
as a different specie). And $\mu_{i} = B_{i}\mu_{B} + S_{i}\mu_{S} + Q_{i}\mu_{Q}$,
where $\mu_{B}$, $\mu_{S}$, and  $\mu_{Q}$ are overall baryon number and strangeness
chemical potentials respectively.

To get particle yields one should consider also entropy density $s$
\begin{equation}\label{entropy}
s={1 \over {6\pi^{2}T^{2}} } \sum_{i=1}^{l} (2s_{i}+1) \int\limits_{0}^{\infty}dp\,
{ {p^{4}} \over { E_{i} } } { { (E_{i} - \mu_{i}) \exp \left\{ {{ E_{i} - \mu_{i} }
\over T} \right\} } \over { \left( \exp \left\{ {{ E_{i} - \mu_{i} } \over T}
\right\} + g_{i} \right)^{2} } }\ .
\end{equation}

To obtain the time dependence of temperature and baryon number and strangeness
chemical potentials one has to solve numerically equations
\eqref{eqstate}-\eqref{entropy} with $s$, $n_{B}$, $n_Q$, and $n_{S}$ given as time
dependent quantities. For $s(t)$, $n_{B}(t)$, and $n_Q(t)$ one obtains expressions
form hydrodynamical calculations and $n_{S}=0$ since we put the overall strangeness
equal to zero during all the evolution.

These equations, enriched by unstable particles effects, form a basic for successful
calculations \cite{brs} of relativistic heavy ion production processes concerning
particle yields and rates. All calculated observables are here the first moments of
related probability distributions. If we are going to get correlations and
fluctuations predictions, we have to calculate second moments. This gives quite new
effects, statistical ensemble dependent,  as was shown in  previous sections devoted
to standard statistical physics approach.

\subsection{Statistical ensembles of high energy physics}

To make our considerations the simplest possible we consider the statistical model
of a non--interacting  gas constrained by the conservation of the abelian charge
$Q$. The thermodynamic system of volume $V$ and temperature $T$ is considered to be
composed of charged particles and their antiparticles carrying charge $\pm 1$
respectively. The requirement of charge conservation in the system is imposed on the
grand canonical or canonical  level. The canonical level means the global charge
conservation while the grand canonical level means the charge conservation on the
average. The partition function of  the above canonical and grand canonical
statistical system is found to be
\begin{subequations}
  \begin{align}\label{part funct}
\mathcal{Z}_Q^{C}(V,T)&=\text{Tr}_Q\,\text{e}^{-\beta\hat{H}}=
\sum\limits_{N_+-N_-=Q}^\infty\frac{z^{N_-+N_+}}{N_-!N_+!} = I_Q(2z)\,,\\
\mathcal{Z}^{GC}(V,T)&=\text{Tr}\,\text{e}^{-\beta(\hat{H} -\mu
\hat{Q})}=\exp{\left(2z\cosh{\frac{\mu}{T}}\right)}\,.
\end{align}
\end{subequations}
where $z$ is the sum over all  one-particle partition functions
{\small
\begin{equation}
  z(T)=\ \frac{V}{(2\pi)^3}\sum_i g_i\int d^3p\,e^{-\beta\sqrt{p^2 +m_i^2}}
  =\ \frac{V}{2\pi^2}T\sum_i g_i\,m_i^2\,K_2\left(\frac{m_i}{T}\right)\equiv V z_0(T)\,,
\end{equation}
} and $g_i$ is the spin degeneracy factor. The sum is taken over all charged
particles and resonances of mass $m_i$ carrying the charge $\pm 1$. The functions
$I_Q$ and $K_2$ are modified Bessel functions. The chemical potential $\mu$
determines the average charge in the grand canonical ensemble
\[\langle Q\rangle=T\frac{\partial}{\partial\mu}\ln\mathcal{Z}^{GC}\,.\]%
This allows to eliminate the chemical potential from further formulae for the grand
canonical probabilities distributions
\begin{equation}\label{chem potQ}
    \frac{\mu}{T}=\text{arcsinh}\frac{\langle Q\rangle}{2 z} =
    \ln\frac{\langle Q\rangle + \sqrt{\langle Q\rangle^2+4 z^2}}{2
    z}\,.
\end{equation}

In the canonical ensemble we have a system of volume $V$ and  total charge $Q$. In
the grand canonical ensemble we have a system with volume $V$ and average charge
$\langle Q\rangle $. Number of particles is not conserved in both ensembles. Number
$N_-$ of negative charged particles shall be extracted from the relevant probability
distributions. The probability distribution $\mathcal{P}_Q^{C}(N_-,V)$ to have $N_-$
negatively  and $N_+=N_-+Q$ positively charged particles is obtained \cite{crt1,kkl}
from the partition function \eqref{part funct} as
 \begin{equation}\label{prob C N}
    \mathcal{P}_Q^{C}(N_-,V)= \frac{z^{2N_-+Q}}{N_-!(N_-+Q)!}\frac{1}{I_Q(2 z)}\,.
\end{equation}
On the other hand  in the GC ensemble with volume $V$ and  average charge $\langle
Q\rangle $ the probability distribution $\mathcal{P}_{\langle
Q\rangle}^{GC}(N_-,Q,V)$ to find a system with a given charge $Q$ and a given number
of negatively charged particles $N_-$ is expressed \cite{crt1} as the product
\begin{equation}\label{prob GC NQ}
\mathcal{P}_{\langle Q\rangle}^{GC}(N_-,Q,V) =
\mathcal{P}_Q^{C}(N_-,V)\,\mathcal{P}_{\langle Q\rangle}^{GC}(Q,V)\,,
\end{equation}
of the canonical particle number distribution $\mathcal{P}_Q^{C}(N_-,V)$ from Eq.
\eqref{prob C N} and the grand canonical probability distribution
\begin{equation}\label{prob GC Q}
    \mathcal{P}_{\langle Q\rangle}^{GC}(Q,V)=
  I_Q(2z)\left[\frac{\langle Q\rangle +
  \sqrt{\langle Q\rangle^2+4 z^2}}{2 z}\right]^Q \text{e}^{-\sqrt{\langle Q\rangle^2+4  z^2}}\,
\end{equation}
to find the total charge $Q$ in the system with the average charge $\langle
Q\rangle$.

\subsection{The thermodynamic limit}
The thermodynamic limit is understood as a limit $V\to\infty$ such that densities of
the system remain constant. So we have for the canonical ensemble
\[Q\to\infty,\ N_-\to\infty\,;\ \frac{Q}{V}=q\,;\ \frac{N_-}{V}=n_-\,\]%
and
\[\langle Q\rangle\to\infty,\ N_-\to\infty\,;\ \frac{\langle Q\rangle}{V}=\langle q\rangle\,;\
 \frac{N_-}{V}=n_-\]%
for the grand canonical ensemble.

To formulate correctly  the thermodynamic limit of quantities involving densities,
one defines  the following probabilities
\begin{subequations}
  \begin{eqnarray}
{\mathbf{P}}_{q}^{C}(n_-,V) &:= & V\mathcal{P}_{V q}^{C}(V
n_-,V)\,,\\
{\mathbf{P}}_{\langle  q\rangle}^{GC}(n_-,q,V)&:= &
V^2\mathcal{P}_{V\langle q\rangle}^{GC}(V n_-,V q,V)\,,\\
{\mathbf{P}}_{\langle  q\rangle}^{GC}(q,V)&:= & V\mathcal{P}_{V\langle
q\rangle}^{GC}(V q,,V)\,.\label{prob GC dens}
 \end{eqnarray}
\end{subequations}

We are going to proceed now in a similar way as in the former section. In a large
volume limit one gets
  \begin{equation}\label{prob corrts C}
    {\mathbf{P}}^{C}_q(n_-,V) =
    \mathcal{P}_{q}^{\infty}(n_-)+\frac{1}{V}R^{C}_q(n_-)+\mathcal{O}(V^{-2})\,,
   \end{equation}
\begin{subequations}\label{prob corrts}
  \begin{eqnarray}
    {\mathbf{P}}^{GC}_{\langle  q\rangle}(n_-,q,V) &=&
    \mathcal{P}_{\langle  q\rangle}^{\infty}(n_-,q)+\frac{1}{V}R^{GC}_{\langle q\rangle}(n_-,q) +
    \mathcal{O}(V^{-2})\,,\label{prob corrts GC nq}\\
    {\mathbf{P}}^{GC}_{\langle  q\rangle}(q,V) &=&
    \mathcal{P}_{\langle  q\rangle}^{\infty}(q)+\frac{1}{V}S^{GC}_{\langle q\rangle}(q) +
    \mathcal{O}(V^{-2})\,.\label{prob corrts GC q}
\end{eqnarray}
\end{subequations}
All functional coefficients can be obtained here using Laplace-Watson theorem. From
the careful analysis one gets \cite{crt2} for the probability distribution
\eqref{prob corrts C} of the canonical ensemble
\begin{equation}\label{TC_NLO}
\begin{split}
  {\mathbf{P}}^{C}_q(n_-;V) =\
      &\delta\left(n_--\langle n_-\rangle_{\infty}\right) +
      \frac{1}{V}\frac{z_0^2}{q^2+4z_0^2}\,\delta'\left(n_--\langle n_-\rangle_{\infty}\right)\\ &+
      \frac{1}{V}\frac{z_0^2}{2\sqrt{q^2+4z_0^2}}\,\delta''\left(n_--\langle
      n_-\rangle_{\infty}\right) + \mathcal{O}(1/V^2)\,,
\end{split}
      \end{equation}
and for the probabilities distributions \eqref{prob corrts} of the grand canonical
distribution
\begin{subequations}\label{TGC_NLO}
\begin{equation}
\begin{split}
  {\mathbf{P}}^{GC}_{\langle  q\rangle}( q,n_-;V) =\ &
    \delta\left(n_--\langle n_-\rangle_{\infty}\right)\delta(q-\langle q\rangle)\\ +
    &\frac{\langle n_-\rangle_{\infty}}{2V}
    \delta''\left(n_--\langle n_-\rangle_{\infty}\right)\delta(q-\langle q\rangle) + \mathcal{O}(1/V^2)\,,
\end{split}
\end{equation}
\begin{equation}
{\mathbf{P}}^{GC}_{\langle  q\rangle}(q,V)=
  \delta\left( q-\langle q\rangle\right) +
  \frac{\sqrt{\langle q\rangle^2+4z_0^2}}{2V  }\,\delta^{''}( q -\langle q\rangle) + \mathcal{O}(1/V^2)\,,
\end{equation}
\begin{equation}
{\mathbf{P}}^{GC}_{\langle  q\rangle}(n_-,V)=
    \delta\left(n_--\langle n_-\rangle_{\infty}\right) +
    \frac{\langle n_-\rangle_{\infty}}{2V}
    \delta''\left(n_--\langle n_-\rangle_{\infty}\right) + \mathcal{O}(1/V^2)\,.
\end{equation}
\end{subequations}
An average limiting  density of charged particles
\begin{equation}\label{dens part pm}
    \langle n_\pm\rangle_{\infty}= \frac{\sqrt{q^2+4z_0^2} \pm q}{2}
\end{equation}
is used in above formulae.
\subsection{Particle moments}
Probability distributions \eqref{TC_NLO} and \eqref{TGC_NLO} allow to write compact
expressions for for particle and charge distribution density moments of any order up
to $\mathcal{O}(1/V^2)$ terms. For particle moments one gets
\begin{equation}\label{TC moments_NLO}
\langle n_\pm^k\rangle^C\simeq\langle n_\pm\rangle_\infty^k
      -\frac{k}{V}\frac{z_0^2}{q^2+4z_0^2}\langle n_\pm\rangle_\infty^{k-1}+
    \frac{k(k-1)}{2V}\frac{z_0^2}{\sqrt{q^2+4z_0^2}}\langle
    n_\pm\rangle_\infty^{k-2}\,,
\end{equation}
for the canonical ensemble moments and
\begin{equation}
      _\langle {n_\pm^k}\rangle^{GC}\simeq\langle n_\pm\rangle_{\infty}^k +
      \frac{k(k-1)}{2V}
      \langle n_\pm\rangle_{\infty}^{k-1}\,.
\end{equation}
for the grand canonical ensemble moments.

Although those moments are density moments they can be expressed by directly
observable variables. Using Eq \eqref{dens part pm} one gets moments as functions of
$q/z_0$ ratio. This ratio is observable as it can be written as a function of the
ratio of charged particles
\[\frac{q^2}{z_0^2}=\frac{\langle N_+\rangle_\infty}{\langle N_-\rangle_\infty} +
\frac{\langle N_-\rangle_\infty}{\langle N_+\rangle_\infty} - 2\,.\]%

\subsection{Semi-intensive variables}
Now we are in position to create some semi-intensive variables. They are finite in
T-limit and have different values dependently on how the charge conservation is
implemented in the description of the  system. There is  actually  a broad  class of
variables. We take an an example
  \begin{equation}\label{particle mom sk}
    \mathcal{S}_k=\frac{\langle N^k\rangle - \langle N\rangle^k}{\langle N\rangle^{k-1}}
\end{equation}

 Indeed from \eqref{TC moments_NLO} and \eqref{TGC_NLO} one gets canonical and grand
 canonical values for positive(negative) particles in the thermodynamic limit
 (denoted as T-limit in the subsequent formulae)
\begin{subequations}
  \begin{equation}\label{TC sk}
    \text{T-}\lim\mathcal{S}_k^{C}=
    \frac{k(k-1)}{4}\frac{\sqrt{q^2+4z_0^2}\mp q}{\sqrt{q^2+4z_0^2}}\,,
\end{equation}
for the canonical ensemble, while in the grand canonical ensemble
\begin{equation}\label{TGC sk}
    \text{T-}\lim\mathcal{S}_k^{GC}
   =\frac{k(k-1)}{2}\,.
\end{equation}
\end{subequations}
The scaled variance is just a  special case of $S_k$ corresponding to $k=2$.

Another examples are classes of variables closely related to cumulant or factorial
cumulant moments \cite{crt2} or susceptibility ratios. Let define $p-$th order
susceptibility
\[\kappa_p=\frac{\partial^p\ln{\mathcal{Z}}}{\partial\mu^p}\,.\]%
One can easily check that the ratios
\begin{equation}
    \mathcal{K}_{p;r}=\frac{\kappa_p}{\kappa_r}\,,
\end{equation}
are semi--intensive quantities.

One can also construct more involved semi-inclusive variables having a finite
T-limit behavior which are determined by higher order asymptotic terms of the
corresponding probability distributions.

\section{Conclusion}
 We have discussed  the differences in the asymptotic properties of the probability
functions for a system with an exact, that is canonical, and with an average, that
is grand canonical,  implementation of  charge conservation. We have shown that in
the thermodynamic limit the corresponding probability distributions in the grand
canonical and canonical ensembles coincide and are described as  generalized
functions. This property is a direct consequence of the grand canonical and
canonical ensemble equivalence in the thermodynamic limit. However, the first finite
volume corrections to the asymptotic value  differ for both ensembles.

Finally, using the results of the probability functions we have derived the
asymptotic behavior  of the charged particle moments and established the differences
in the grand canonical and canonical formulation. We have also applied these results
to find the thermodynamic limit of a class of  semi--intensive quantities. It was
shown that in systems with exact and average charge conservation such  quantities
should naturally converge to different values in the thermodynamic limit. This is
because the behavior  of the semi--intensive quantities in the near vicinity to the
thermodynamic limit are determined by the subleading, finite volume, corrections to
the probability distributions which are specific to a given statistical ensemble.

Is important that first moments are the same in the canonical and grand canonical
ensemble. This means that particle yields in heavy ion collision and equation of
state of dense hadronic medium are insensitive to the statistical ensemble in the
thermodynamic limit. This is not the case, however, for fluctuations and higher
moments. Finite volume effect are more and more relevant for higher moments. Such a
situation appears when comparing the statistical model with lattice gauge theory
results obtained on a small lattice.

\end{document}